\documentclass[12pt]{article}
\usepackage{pgf}
\usepackage{amsmath,amssymb,bbm}
\usepackage{latexsym}
\usepackage{caption}
\usepackage{amsmath}
\usepackage{amstext}
\usepackage{amsthm}
\usepackage{amssymb}
\usepackage{amsopn}
\usepackage{mathrsfs} %\mathscr
\usepackage{dsfont} %\mathds
\usepackage[bottom]{footmisc}
\usepackage{indentfirst}
\usepackage{upref}
\usepackage{cite}
\usepackage{units}
\usepackage{fancybox}
\usepackage%[lflt]
{floatflt}
\numberwithin{equation}{section}

\renewcommand{\baselinestretch}{1.409}
\begin{document}
\begin{titlepage} 
\renewcommand{\baselinestretch}{1.1}
\small\normalsize
\begin{flushright}
%arXiv/\\
MZ-TH/08-19
\end{flushright}

\vspace{0.1cm}
\vspace{0.1cm}
\vspace{0.1cm}
\vspace{0.1cm}
\vspace{1cm}\vspace{1cm}\vspace{1cm}\vspace{0.5cm}
\begin{center}   

{\Large \textbf{Effective Potential of the Conformal Factor:  \\Gravitational Average Action and  \\Dynamical Triangulations %\\[2mm] \\[6mm] 
}}

\vspace{1.4cm}
{\large J.-E.~Daum and M.~Reuter}\\

\vspace{0.7cm}
\noindent
\textit{Institute of Physics, University of Mainz\\
Staudingerweg 7, D--55099 Mainz, Germany}\\

\end{center}

\vspace*{0.6cm}
\begin{abstract}
\vspace{12pt}
We discuss the effective potential of the conformal factor in the effective average action approach to Quantum Einstein Gravity. Without invoking any truncation or other approximations we show that if the theory has has a non-Gaussian ultraviolet fixed point and is asymptotically safe the potential has a characteristic behavior near the origin. We argue that this characteristic behavior has already been observed in numerical simulations within the Causal Dynamical Triangulation approach.
\end{abstract}
\end{titlepage}
%
%
%
%%%%%%%%%%%%%%%%%%%%%%%%%%%%%%%%%%%%%%%%%%%%%%%%%%%%%%%%%%%%
%%%%%%%%%%%%%%%%%%%%%%%%%%%%%%%%%%%%%%%%%%%%%%%%%%%%%%%%%%%%
%
%
%
\section{Introduction}\label{s1}
One of the problems one faces when one tries to construct a fundamental theory of quantum gravity is the absence of any experimental data or observations that necessarily would have to be explained by the theory \cite{kiefer,A,R,T}. In this situation it is particularly important to compare the predictions of a priori different candidate theories and to find out whether they are perhaps just different formulations of the same underlying theory or whether they really belong to different ``universality classes''. In this way one can at least narrow down the set of independent possibilities among which the experiment must decide in the end. Guided by the experience with Yang-Mills theory we would expect that in particular the comparison of continuum and lattice approaches should be very instructive and fruitful.  
On the side of the continuum approaches, recently a lot of efforts went into the exploration of the asymptotic safety scenario \cite{wein,mr,percadou,oliver1,frank1,oliver2,oliver3,oliver4,souma,frank2,prop,oliverbook,perper1,codello,litimgrav,frankmach,creh1,creh2,oliverfrac,jan1,jan2,neuge,max,livrev} in the formulation based upon the gravitational average action. It aims at defining a microscopic quantum field theory of gravity in terms of a complete, i.e., infinitely extended renormalization group (RG) trajectory on the theory space of diffeomorphism invariant functionals of the metric. The limit of an infinite ultraviolet (UV) cutoff is taken by arranging this trajectory to approach a non-Gaussian fixed point (NGFP) at large scales ($k \rightarrow \infty$). This NGFP of the effective average action is not only instrumental in constructing the quantum field theory by dictating how all generalized couplings must ``run'' when the UV regulator scale is sent to infinity, it also determines the physical properties of the resulting regulator-free theory at large {\it physical} scales, the behavior of propagators at large momenta, for instance. We refer to this quantum field theory of the metric, defined in the continuum by means of the effective average action, as Quantum Einstein Gravity (QEG). 

In Yang-Mills theory, the asymptotic scaling behavior that controls the approach to the continuum of the corresponding lattice gauge theory can be read off from the corresponding effective average action \cite{avact,ym,avactrev,ymrev}. It is therefore plausible to ask whether there is an analogous relationship between QEG and gravity on a (dynamical) lattice. A first comparison (of critical indices) has been attempted in \cite{frank1} where the results from the average action were confronted with Monte Carlo data obtained in Regge calculus simulations \cite{hamwill}. The results were found to be consistent, albeit within considerable error bars. 

A different statistical-mechanics based framework for quantum gravity that has been developed during the past years is the Causal Dynamical Triangulation (CDT) approach \cite{ajl1,ajl2,ajl34,agjl}. It provides a background-independent, nonperturbative definition of quantum gravity in which the causal (i.e., Lorentzian) structure of spacetime plays a crucial role. Only causally well-behaved geometries contribute to the path integral. It is regularized by summing over a particular class of triangulated, piecewise flat geometries. All contributing spacetimes are foliated by a kind of proper-time and can be obtained by gluing together 4-simplices in a way that respects this foliation. The slices of constant proper-time consist of spacelike tetrahedra forming a 3-dimensional piecewise flat manifold. Typically its topology is assumed to be that of a three-sphere $S^3$. Each configuration contributing to the regularized path integral can be rotated to Euclidean signature, and the rotated partition function can be computed by means of standard Monte Carlo methods. It was found \cite{agjl} that the CDT approach can generate almost classical 4-dimensional universes on which only small quantum fluctuations are superimposed.

In ref.\ \cite{ajl2} an effective action for the Robertson-Walker scale factor $a (t)$ of these universes has been extracted from the Monte Carlo data. In the present paper we are going to compare this effective action to the corresponding prediction of QEG. We shall demonstrate that the latter reflects the non-Gaussian fixed point, the crucial prerequisite of asymptotic safety, by a characteristic ``finger print'' at small scale factors. The very same finger print has actually been observed already in the CDT simulations.

We interprete these results as an indication pointing in the direction that QEG and the CDT approach could possibly describe the same underlying theory in different languages, i.e. that they are two representatives of the same ``universality class''. 

In fact, in ref.\ \cite{oliver1,oliverfrac} another point of contact between CDT and QEG had been found already: They agree on the microscopic spectral dimension of macroscopically 4-dimensional spacetimes; in either case one finds the somewhat surprising result $d_{\rm S} = 2$. 

The remaining sections of this paper are organized as follows. In Section \ref{s2} we analyze the effective potential of the conformal factor in QEG and discuss in particular the impact the NGFP has on it. In Section \ref{s3} we make contact with the CDT simulations, and in Section \ref{s4} we present the conclusions. 

%%%%%%%%%%%%%%%%%%%%%%%%%%%%%%%%%%%%%%%%%%%%%%%%%%%%%%%%%%%%
%%%%%%%%%%%%%%%%%%%%%%%%%%%%%%%%%%%%%%%%%%%%%%%%%%%%%%%%%%%%
%
%
%
\section{The Effective Potential of the Conformal Factor \\in QEG}\label{s2}
In this section we discuss the standard effective potential (i.\,e., the one with vanishing infrared cutoff, $k = 0$) for the conformal factor of metrics on maximally symmetric spacetimes with the topology of a $d$-dimensional sphere $S^d$. The starting point is the {\it exact} gravitational effective average action \cite{mr} along some RG trajectory, $\Gamma_k [g_{\mu\nu}, \bar{g}_{\mu\nu}]$, and the related reduced functional $\bar{\Gamma}_k [g_{\mu\nu}] \equiv \Gamma_k [g_{\mu\nu}, g_{\mu\nu}]$. (The ghost arguments are set to zero and are not indicated explicitly.) The latter functional is assumed to have a representation of the form
\begin{eqnarray} \label{Gamma} \bar{\Gamma}_k [g_{\mu\nu}] = \sum_{\alpha} \bar{u}_\alpha (k) \: I_\alpha [g_{\mu\nu}] 
\end{eqnarray}
where $\{I_\alpha [g_{\mu\nu}]\}$ is an infinite set of local and nonlocal ``basis'' functionals, invariant under diffeomorphisms acting on $g_{\mu\nu}$, and the $\bar{u}_\alpha$'s are the corresponding running coupling constants. We denote their canonical mass dimensions by $d_\alpha \equiv [\bar{u}_\alpha]$. Hence, since $\bar{\Gamma}_k$ is dimensionless, $[I_\alpha] = - d_\alpha$. The dimensionless running couplings are defined by
\begin{eqnarray} u_\alpha (k) \equiv k^{- d_\alpha}\: \bar{u}_\alpha (k) \nonumber\end{eqnarray}
so that we may rewrite \eqref{Gamma} as
\begin{eqnarray} \label{Gamma-dimless} \bar{\Gamma}_k [g_{\mu\nu}] = \sum_{\alpha} u_\alpha (k)\: k^{d_\alpha} I_\alpha [g_{\mu\nu}] 
\end{eqnarray}

Up to now the metric argument $g_{\mu\nu}$ was completely general. At this point we specialize for metrics on $S^d$, with a variable radius $\phi$. We parametrize them as
\begin{eqnarray} \label{conf-metric} g_{\mu\nu} = \phi^2 \:\hat{g}_{\mu\nu}
\end{eqnarray}
where $\hat{g}_{\mu\nu}$ is the metric on the round $S^d$ with unit radius, and the conformal factor $\phi$ is position independent. Hence $g_{\mu\nu}$ is a metric on a round sphere with radius $\phi$. We shall denote the volume of the unit-$S^d$ by $\sigma_d \equiv \int {\rm d}^d x \: \sqrt{\hat{g}} = 2 \pi^{(d+1)/2} / \Gamma \big( (d+1)/2 \big)$.

We use conventions such that the coordinates $x^\mu$ are dimensionless and $\phi$ has the dimension of a length. Hence $[g_{\mu\nu}] = - 2$, and $\hat{g}_{\mu\nu}$ is dimensionless, $[\hat{g}_{\mu\nu}] = 0$.

Without having made any approximation so far, the effective average potential for the conformal factor, $U_k (\phi)$, by definition, obtains by inserting the special argument \eqref{conf-metric} into $\bar{\Gamma}_k$:
\begin{eqnarray} \label{eff-av-pot} U_k (\phi) \int {\rm d}^d x \: \sqrt{\hat{g}}\: \equiv \: \bar{\Gamma}_k [g_{\mu\nu} = \phi^2 \hat{g}_{\mu\nu}]
\end{eqnarray}
In terms of the expansion \eqref{Gamma-dimless} we have the exact representation \begin{eqnarray} \label{eff-av-pot-conf-1} U_k (\phi) = \sigma_d^{-1} \sum_{\alpha} u_\alpha (k)\: k^{d_\alpha} I_\alpha [\phi^2 \hat{g}_{\mu\nu}]    
\end{eqnarray}
or, more explicitly, 
\begin{eqnarray} \label{eff-av-pot-conf-2} U_k (\phi) = \sigma_d^{-1} \sum_{\alpha} u_\alpha (k)\: (k \phi)^{d_\alpha}\: I_\alpha [\hat{g}_{\mu\nu}]    
\end{eqnarray}
To obtain equation \eqref{eff-av-pot-conf-2} we exploited that $I_\alpha [\phi^2 \hat{g}_{\mu\nu}] = \phi^{d_\alpha} I_\alpha [\hat{g}_{\mu\nu}]$ which holds true since $I_\alpha$ has dimension $- d_\alpha$. (This relation can be regarded the definition of the canonical mass dimension.)  

Eq.\ \eqref{eff-av-pot-conf-2} makes it manifest that if we know a complete RG trajectory $\{ u_\alpha (k), 0 \le k < \infty\}$ we can deduce the exact running potential from it, and in particular its $k \rightarrow 0$ limit, the standard effective potential $U_{\rm eff} (\phi) \equiv U_{k = 0} (\phi)$. Usually we are not in the comfortable situation of knowing trajectories exactly; nevertheless certain important properties of $U_{\rm eff} (\phi)$ can be deduced on general grounds. For this purpose we shall employ the following decoupling argument which is standard in the average action context \cite{avact, avactrev}. 

The basic observation is that the true, i.\,e. dimensionful coupling constants $\bar{u}_\alpha (k)$ have a significant running with $k$ only as long as the number of field modes integrated out actually depends on $k$. If there are competing physical cutoff scales such as masses or field amplitudes the running with $k$ stops once $k$ becomes smaller than the physical cutoff scales. (See Appendix C.3 of \cite{livrev} for an example.) In the case at hand this situation is realized in a particularly transparent way. The quantum metric is expanded in terms of eigenfunctions of the covariant (tensor) Laplacian $\bar{D}^2$ of the metric $\bar{g}_{\mu\nu}$. This metric corresponds to a sphere of radius $\phi$; hence all eigenvalues of the Laplacian are discrete multiples of $1/{\phi^2}$. As a result, when $k$ has become as small as $k \approx 1/\phi$, the bulk of eigenvalues is integrated out, and the $\bar{u}_\alpha$'s no longer change much when $k$ is lowered even further. Therefore we can approximate
\begin{eqnarray} \label{approx} U_{\rm eff} (\phi) \:\equiv\: U_{k=0} (\phi) \approx U_{k = 1/\phi} (\phi)\end{eqnarray}

In order to make the approximation \eqref{approx} strictly valid we have to be slightly more specific about the precise definition of $U_k (\phi)$. The above argument could be spoiled by zero modes of $\bar{D}^2$. Therefore we define $\Gamma_k$ and $U_k$ in terms of a functional integral over the fluctuation modes of the metric with a non-zero eigenvalue of $\bar{D}^2$ only. As a result, the actual partition function would obtain by a final integration over the zero modes which is not performed here. The only zero mode relevant in the case at hand is the fluctuation of the conformal factor. When we compare our results to the CDT simulations later on it will be important to keep in mind that $U_{\rm eff} (\phi)$ has the interpretation of an effective potential in which the conformal fluctuation has not yet been integrated out since it is essentially this quantity that is provided by the Monte Carlo simulations.
  
Eq.\ \eqref{approx} has a simple intuitive interpretation in terms of coarse graining: By lowering $k$ below $1/\phi$ one tries to ``average'' field configurations over a volume that would be larger than the volume of the whole universe. As this is not possible, the running stops. Note that the $S^d$ topology enters here; the finite volume of the sphere is crucial. 

With the approximation \eqref{approx} we obtain the following two equivalent representations of $U_{\rm eff} (\phi)$ in terms of the dimensionless and dimensionful running couplings, respectively:
\begin{eqnarray} \label{eff-pot-dimless} U_{\rm eff} (\phi) &=& \sigma_d^{-1} \sum_{\alpha} u_\alpha (\phi^{-1}) \:I_\alpha [\hat{g}_{\mu\nu}] \\
 \label{eff-pot-dimful} U_{\rm eff} (\phi) &=& \sigma_d^{-1} \sum_{\alpha} \bar{u}_\alpha (\phi^{-1}) \:\phi^{d_\alpha} I_\alpha [\hat{g}_{\mu\nu}]   
\end{eqnarray}
As an application of these representations we consider two special cases.

Let us assume the RG trajectory under consideration has a {\it classical regime} between the scales $k_1$ and $k_2$, meaning that $\bar{u}_\alpha (k) \approx {\rm const} \equiv \bar{u}_\alpha^{\rm class}$ for $k_1 < k < k_2$. Then \eqref{eff-pot-dimful} implies that for $k_2^{-1} < \phi < k_1^{-1}$, approximately, 
\begin{eqnarray} \label{eff-pot-class} U_{\rm eff} (\phi) &=& \sigma_d^{-1} \sum_{\alpha} \bar{u}_\alpha^{\rm class} \:I_\alpha [\hat{g}_{\mu\nu}] \:\phi^{d_\alpha}\end{eqnarray} 
As expected, this potential has a nontrivial $\phi$-dependence governed by the classical couplings $\bar{u}_\alpha^{\rm class}$. 

Next let us explore the consequences which a non-Gaussian fixed point has for the effective potential. We assume that the dimensionless couplings $u_\alpha (k)$ approach fixed point values $u_\alpha^\ast$ for $k \rightarrow \infty$. More precisely, we make the approximation $u_\alpha (k) \approx u_\alpha^\ast$ for $ k \gtrsim M$ with $M$ the lower boundary of the asymptotic scaling regime. Then the representation \eqref{eff-pot-dimless} tells us that $U_{\rm eff} (\phi) = \sigma_d^{-1} \sum_{\alpha} u_\alpha^\ast\:I_\alpha [\hat{g}_{\mu\nu}]$ if $\phi \lesssim M^{-1}$. Obviously this potential is completely independent of $\phi$:
\begin{eqnarray} \label{eff-pot-fp} U_{\rm eff} (\phi) = {\rm const} \hspace{0.3cm}\mbox{for all}\:\:\: \phi \lesssim M^{-1}
\end{eqnarray}
In typical applications (see below), $M$ equals the Planck mass $m_{\rm Pl} \equiv \ell^{-1}_{\rm Pl}$ so that $U_{\rm eff}$ is constant for $\phi \lesssim \ell_{\rm Pl}$.

Eq.\ \eqref{eff-pot-fp} is one of our main results. It shows that the existence of an ultraviolet fixed point has a characteristic impact on the effective potential of the conformal factor: Regardless of all details of the RG trajectory, the potential is completely flat for small $\phi$. The interpretation of this result is that for $\phi \lesssim M^{-1}$ the cost of energy (Euclidean action) of a sphere with radius $\phi$ does not depend on $\phi$. Spheres of any radius smaller than $M^{-1}$ are on an equal footing. This is exactly the kind of fractal-like behavior and scale invariance one would expect near the NGFP \cite{oliver1,oliver2}.

We emphasize that except for the decoupling relation \eqref{approx} no approximation went into the derivation of this result. It is an exact consequence of the assumed asymptotic safety, the existence of a NGFP governing the short distance behavior. Neither has the theory space been truncated nor have any fields been excluded from the quantization (such as in conformally reduced gravity the transverse tensors, for instance, cf.\ \cite{creh1,creh2}). 

On the basis of the above general argument we cannot predict how precisely, or how quickly the effective potential flattens when we approach the origin. However, we expect that its derivative $\partial U_{\rm eff}/{\partial \phi}$ vanishes at $\phi = 0$. This has an important physical implication. In general, possible vacuum states of the system (the ``universe'') can be found from the effective field equation $\delta \Gamma_{k = 0}/{\delta g_{\mu\nu}} = 0$. More specifically, $S^d$-type groundstate candidates have a radius $\phi_0$ at which $\big(\partial U_{\rm eff}/{\partial \phi}\big) (\phi_0) = 0$. Thus we see that thanks to the NGFP a vanishing radius $\phi_0 = 0$ has become a vacuum candidate, the derivative of $ U_{\rm eff}$ vanishes there. (To qualify as the true vacuum it should be the global minimum.) Hence the universe has an at least metastable stationary state with $\phi = 0$, i.e. a state with a vanishing metric expectation value $\langle g_{\mu\nu} \rangle = 0$. In this state gravity is in a {\it phase of unbroken diffeomorphism invariance}, see ref. \cite{creh2} for a discussion of this phase in the context of asymptotic safety.

Let us finally illustrate the above discussion in the familiar setting of the Einstein-Hilbert truncation \cite{mr} in $d = 4$ which is defined by the ansatz
\begin{eqnarray} \label{e-h-trunc} \bar{\Gamma}_k [g_{\mu\nu}] = - \frac{1}{16 \pi G_k} \int{\rm d}^4 x \: \sqrt{g}\Big( R(g) - 2 \Lambda_k \Big)      
\end{eqnarray}
Inserting (\ref{conf-metric}) with $\phi = \phi (x)$ we obtain
\begin{eqnarray} \label{e-h-trunc-conf} \bar{\Gamma}_k [\phi^2 \hat{g}_{\mu\nu}] = \frac{3}{4 \pi G (k)} \int{\rm d}^4 x \: \sqrt{\hat{g}} \Big[- \frac{1}{2}\hat{g}^{\mu\nu} \partial_\mu \phi \partial_\nu \phi  - \phi^2 + \frac{1}{6} \Lambda (k)\phi^4 \Big]     
\end{eqnarray}
For $x$-independent $\phi$ only the potential term survives, with
\begin{eqnarray} \label{e-h-trunc-conf-pot} U_k (\phi) &=& \frac{3}{4 \pi G (k)} \Big(- \phi^2 + \frac{1}{6} \Lambda (k)\:\phi^4 \Big)\\
&=& \frac{3}{4 \pi g (k)} \Big(- k^2 \phi^2 + \frac{1}{6} \lambda (k)\: k^4 \phi^4 \Big)\nonumber     
\end{eqnarray}
If $\Lambda (k) > 0$, the case we shall always consider in the following, $U_k (\phi)$ has a minimum at a nonzero radius given by 
\begin{eqnarray} \label{class-minimum} \phi_0 (k) = \sqrt{3/\Lambda (k)}
\end{eqnarray}
This is exactly the radius of the $S^4$ which solves the ordinary Einstein equation following from the action \eqref{e-h-trunc}\footnote{Because this space is maximally symmetric, by Palais' theorem \cite{palais}, inserting the ansatz $g_{\mu\nu} = \phi^2 \hat{g}_{\mu\nu}$ commutes with deriving the critical point.}. In the second line of \eqref{e-h-trunc-conf-pot} we employed the dimensionless Newton constant $g (k) \equiv k^2 G (k)$ and cosmological constant $\lambda (k) \equiv \Lambda (k)/k^2$. So there are the following two equivalent ways of writing the effective potential:
\begin{eqnarray} \label{e-h-trunc-conf-eff-pot-dimful} U_{\rm eff} (\phi) &=& \frac{3}{4 \pi} \Big[- \frac{1}{G (\phi^{-1})}\: \phi^2 + \frac{1}{6} \frac{\Lambda (\phi^{-1})}{G (\phi^{-1})}\:\phi^4 \Big]\\
 \label{e-h-trunc-conf-eff-pot-dimless} U_{\rm eff} (\phi) &=& \frac{3}{4 \pi} \Big[- \frac{1}{g (\phi^{-1})} + \frac{1}{6} \frac{\lambda (\phi^{-1})}{g (\phi^{-1})}\Big]
\end{eqnarray}

The RG trajectories of the Einstein-Hilbert truncation have been investigated and classified in \cite{frank1}. Here we can concentrate on those with a positive cosmological constant, those of ``Type IIIa'', because this class of trajectories seems to be the one which is behind the Monte Carlo results we are going to discuss.
Important regimes along a Type IIIa trajectory include\\
{\bf The NGFP regime:} $g (k) \approx g_\ast$, $\lambda (k) \approx \lambda_\ast$ for $k \gtrsim M$.\\
{\bf The $k^4$ regime:} $G (k) \approx {\rm const}$, $\Lambda (k) \propto k^4$ for $k_{\rm T} \lesssim k \lesssim M$, where $k_{\rm T}$ is the ``turning point'' scale at which $\beta_\lambda$ vanishes.\\ 
{\bf The classical regime:} $G (k) \approx {\rm const} \equiv \bar{G}$, $\Lambda (k) \approx {\rm const} \equiv \bar{\Lambda}$ for $k_{\rm term} \ll k \lesssim k_{\rm T}$ where $k_{\rm term}$ is the scale at which the Einstein-Hilbert truncation breaks down and the trajectory terminates at a singularity\footnote{If one tentatively matches the trajectory against the observed values of $G$ and $\Lambda$ one finds that $k_{\rm T} \approx 10^{-30} m_{\rm Pl}$, corresponding to $k^{-1}_{\rm T} \approx 10^{-3} {\rm cm}$, and  $k_{\rm term} \approx 10^{-60} m_{\rm Pl} \approx H_0$ so that $k^{-1}_{\rm term}$ equals about the present Hubble radius \cite{h3}, \cite{entropy}.}.\\ 
If one defines the classical Planck mass and length by $m_{\rm Pl} \equiv \ell^{-1}_{\rm Pl} \equiv \bar{G}^{-1/2}$ one finds that, approximately, $M \approx m_{\rm Pl}$. (For further details see \cite{frank1,h3,entropy}; see in particular Fig. 4 of \cite{h3}.)

In the $k^4$-regime, when $k$ decreases, the cosmological constant quickly becomes smaller proportional to $k^4$, and the radius of the sphere ``on shell'', $\phi_0 (k)$, increases proportional to $1/k^2$.

If the underlying RG trajectory of QEG is of Type IIIa then $U_{\rm eff} (\phi)$ is constant in the NGFP regime $\phi \lesssim \ell_{\rm Pl}$, and it equals the classical potential for $k^{-1}_{\rm T} \lesssim \phi \lesssim k^{-1}_{\rm term}$. Note that our ignorance about the infrared end of the trajectory entails that we have no information about the effective potential {\it for large values of $\phi$}. The intermediate $k^4$-regime of the trajectory gives rise to a behavior
\begin{eqnarray} \label{e-h-trunc-conf-eff-pot-k4} U_{\rm eff} (\phi) \propto (- \phi^2 + {\rm const})\hspace{0.3cm}\mbox{for}\hspace{0.3cm}k^{-1}_{\rm T} \lesssim \phi \lesssim \ell_{\rm Pl}.
\end{eqnarray}

In the above discussion we tacitly assumed that the trajectory is such that $M^2 \approx m^2_{\rm Pl} \gg \bar{\Lambda}$; otherwise no classical regime would exist.
\begin{figure}[h]
\begin{center}
\includegraphics[width=0.7\textwidth]{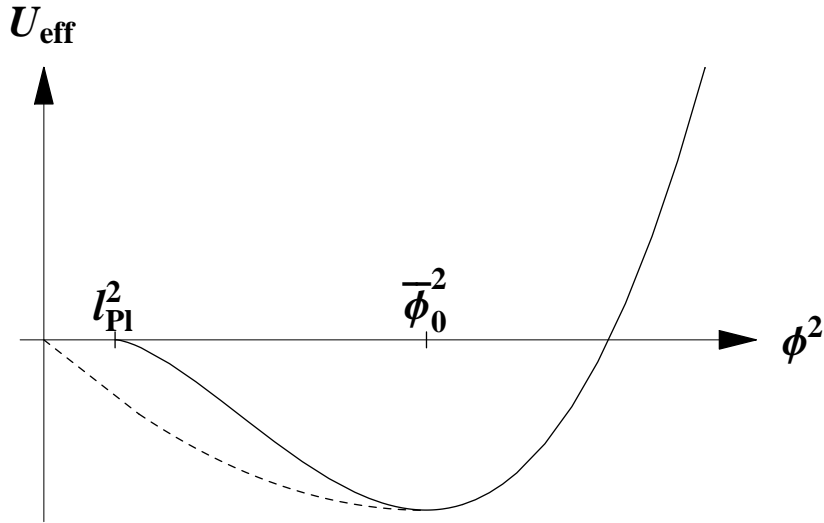}
\caption{The effective potential for the Type IIIa trajectory discussed in the text. The dashed line represents the potential $U_{\rm class}$ with the same values of $\bar{G}$ and $\bar{\Lambda}$, but all quantum effects neglected.}
\label{eff-pot-fig}
\end{center}
\end{figure}

A qualitative sketch of the resulting $U_{\rm eff}$ is shown in Fig.\ \ref{eff-pot-fig}. It is compared there to the classical potential $U_{\rm class}$ which would obtain if $G$ and $\Lambda$ had no $k$-dependence at all. The crucial difference between the two is the almost constant $U_{\rm eff}$ at small $\phi$. This regime is a pure quantum gravity effect, directly related to the existence of a NGFP. Quantum mechanically, but not classically, the universe can be stationary at small values of $\phi$, at least at $\phi = 0$.

As a consequence of our assumption $\bar{\Lambda} \ll m^2_{\rm Pl}$, the $U_{\rm eff} = {\rm const}$ regime ends at a radius $\phi \approx \ell_{\rm Pl}$ which is {\it smaller} than the classical ``on-shell'' radius $\bar{\phi}_0 = \sqrt{3/\bar{\Lambda}}$. The actual ``size of the universe'' corresponds to a scale in the classical regime of the RG trajectory therefore.

In the region where the quantum effects modify $U_{\rm class}$ most strongly the term $\propto \phi^2$ is the dominant one. We can therefore say that the key effect behind the flattening of the potential near the origin is the running of Newton's constant. Its consequence for the shape of $U_{\rm eff}$ can be understood as the result of the ``RG improvement'' \cite{bh,erick1,cosmo1,cosmofrank,cosmo2,entropy,esposito,h1,h2,h3,girelli,mof} 
\begin{eqnarray} \label{Newton-RG-improv} \frac{1}{G} \phi^2 \: \longrightarrow\: \frac{1}{G(k = \phi^{-1})} \phi^2 \:=\: \frac{1}{g_\ast}
\end{eqnarray}
with $G (k) = g_\ast/k^2$, as appropriate near the NGFP.

%
%
%
%%%%%%%%%%%%%%%%%%%%%%%%%%%%%%%%%%%%%%%%%%%%%%%%%%%%%%%%%%%%
%%%%%%%%%%%%%%%%%%%%%%%%%%%%%%%%%%%%%%%%%%%%%%%%%%%%%%%%%%%%
\section{Making Contact with CDT Simulations}\label{s3}
The CDT approach \cite{ajl1,ajl2,ajl34,agjl} defines a discrete version of the Wick rotated quantum-gravitational proper-time propagator 
\begin{eqnarray} \label{CDT-prop} G^{\rm E}_{\Lambda, G} \big[g_3 (0), g_3 (t) \big] = \int{\cal D} g_{\rm E}\: {\rm e}^{- S_{\rm E} [g_{\rm E}]}
\end{eqnarray}
Here $S_{\rm E}$ is the Euclidean Einstein-Hilbert action, and the integration is over all 4-dimensio-\\nal Euclidean geometries $g_{\rm E}$ of topology $S^3 \times [0, 1]$, each with proper-time running from $0$ to $t$, and with prescribed spatial boundary geometries $g_3 (0)$ and $g_3 (t)$, respectively. In the simulations reported in \cite{ajl2}, for technical reasons, periodic rather than fixed boundary conditions have been used so that the topology of the spacetimes summed over is $S^3 \times S^1$ rather than $S^3 \times [0, 1]$. (Furthermore, the simulations were done at constant $4$-volume $V_4$ rather than constant $\Lambda$; the corresponding propagator is related to \eqref{CDT-prop} by a Laplace transformation.) 

It is instructive to visualize the typical, statistically representative $4$-geometries contributing to the path integral. They are characterized by a function $V_3 (s)$, $0 \le s \le t$, where $V_3 (s)$ is the $3$-volume of the spatial $S^3$ at proper-time $s$. If $t$ is large enough, a ``typical universe'' has long epochs with a very small $V_3$ at early and late times (the ``stalk'') and in between a region with a large $V_3 (s)$, see Fig. 1 of ref.\ \cite{ajl2}. 

It has been shown \cite{ajl2} that the dynamics of these ``universes'' is well reproduced by a minisuperspace effective action for rotated Robertson-Walker metrics
\begin{eqnarray} \label{R-W-metric} {\rm d}s^2 = {\rm d}t^2  + a^2 (t)\: {\rm d}\Omega^2_3
\end{eqnarray}
where ${\rm d}\Omega^2_3$ is the line element of the unit $3$-sphere so that $V_3 (s) \propto a^3 (s)$. It reads
\begin{eqnarray} \label{eff-act-minisuper} S_{\rm eff} [a] = - \frac{3 \sigma_3}{8 \pi} \frac{1}{G} \int_{0}^{t}{\rm d}s\Big\{ - a(s) \Big(\frac{{\rm d} a(s)}{{\rm d}s}\Big)^2 + V \big(a(s)\big)\Big\}
\end{eqnarray}
For large $a$ the potential $V$ is 
\begin{eqnarray} \label{eff-pot-minisuper}  V (a) = - a + \frac{1}{3} \Lambda a^3 \equiv V_{\rm cl} (a)
\end{eqnarray}
The action \eqref{eff-act-minisuper} with \eqref{eff-pot-minisuper} is, up to an overall minus sign, what one obtains when one inserts \eqref{R-W-metric} into the Einstein-Hilbert action. (In simulations with fixed $V_4$ the constant $\Lambda$ is a Lagrange multiplier to be fixed such that $\int_{0}^{t}{\rm d}s\:V_3 (s) = V_4$.) 
For small $a$, the $(-a)$-term in \eqref{eff-pot-minisuper} is to be replaced by a function of $V_3$ whose derivative at $a = 0$ behaves as $V_3^\nu$, $\nu \ge 0$. A convenient parametrization, valid for all $a$, is 
\begin{eqnarray} \label{eff-pot-minisuper-small-a} V (a) = 1 - (1+a^3)^{1/3} + \frac{1}{3} \Lambda a^3
\end{eqnarray}

\begin{figure}[h]
\begin{center}
\includegraphics[width=0.7\textwidth]{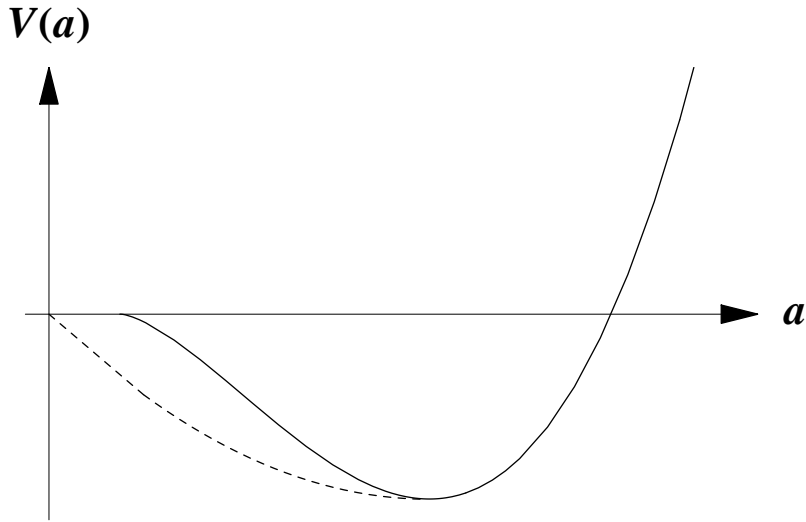}
\caption{The effective potential $V(a)$ of eq.\ \eqref{eff-pot-minisuper-small-a} obtained from the Monte Carlo simulations (solid line). The dashed line shows the potential $V_{\rm cl} (a)$ from the classical Einstein-Hilbert action. (From Loll et al. \cite{ajl2})}
\label{loll-fig}
\end{center}
\end{figure}

This potential is sketched in Fig.\ \ref{loll-fig} and compared to $V_{\rm cl} (a)$. 
While the latter has a negative slope at the origin, $V^\prime_{\rm cl} (0) = -1$, the potential \eqref{eff-pot-minisuper-small-a} with the corrected small-$a$ behavior has vanishing derivative there: $V^\prime (0) = 0$. This quantum induced modification is important since \eqref{eff-pot-minisuper-small-a} allows for a classically stable solution $a (t) = 0$ which explains the stalk observed in the computer simulations \cite{ajl2}.

To summarize: According to the Monte Carlo simulations, the effective potential of the Robertson-Walker scale factor agrees with the classical one for large $a$, but differs from $V_{\rm cl}$ at small $a$ in a characteristic way: $V (a)$ becomes essentially constant (zero) for $ a \rightarrow 0$. In the parametrization \eqref{eff-pot-minisuper-small-a} the classical $V (a) \approx - a$ ist turned into $V (a) \approx - \frac{1}{3} a^3$ by the quantum effects, which amounts to an almost flat $a$-dependence for $a \ll 1$.

Comparing Figs.\ \ref{eff-pot-fig} and \ref{loll-fig} makes it obvious that in both cases the dominant quantum gravity effects have an analogous impact on the effective potentials $U_{\rm eff} (\phi)$ and $V (a)$, respectively: Their slope at the origin , while strictly negative classically, vanishes thanks to the quantum effects, and for small $\phi$, or $a$, the potential is essentially flat, rendering a state with $a = 0$ or $\phi = 0$ stationary. 

It is plausible to assume that we are dealing with the same phenomenon in both cases. In fact, upon introducing the conformal time $\eta (t) = \int^{t} {\rm d} t^\prime\:/a (t^\prime)$ the line element \eqref{R-W-metric} assumes a form analogous to \eqref{conf-metric},
\begin{eqnarray} \label{R-W-metric-conf-time} {\rm d}s^2 = \phi (\eta)^2 \big[{\rm d}\eta^2 + {\rm d}\Omega^2_3\big]
\end{eqnarray}
with the conformal factor $\phi (\eta) \equiv a (t(\eta))$. Since $\phi$ and $a$ differ only by a time reparametrization, which is irrelevant here, the potentials $U_{\rm eff} (\phi)$ and $V (a)$ are almost the same object. In particular we defined $U_{\rm eff} (\phi)$ in terms of a functional integral (or the corresponding flow equation) which does not include the conformal zero mode, i.\,e. fluctuations which merely change the radius of the $S^4$. Likewise its CDT counterpart $V (a)$ results from integrating out all modes other than the spatially constant global scale.

Thus we see that within QEG the non-Gaussian fixed point predicts a vanishing slope of $U_{\rm eff}$ at $\phi = 0$, and upon identifying $U_{\rm eff}$ with $V (a)$ this feature is actually observed in the CDT simulations: They all generate universes with a ``stalk'', and according to the analysis in \cite{ajl2} this stalk owes its existence to the vanishing derivative of $V (a)$ at $a = 0$. In a certain sense, the stalk seems to realize the ``exotic'' phase of gravity when the metric has zero expectation value and the group of diffeomorphisms is not broken by the groundstate (see also section 6 of \cite{creh2}).

Another minor difference between the QEG and CDT setting, respectively, is that $\hat{g}_{\mu\nu}$ is a metric on $S^4$, while ${\rm d}\eta^2 + {\rm d}\Omega^2_3$ refers to $S^3 \times [0, 1]$ or $S^3 \times S^1$. However, we do not expect such global issues to cause qualitative changes for small conformal factors\footnote{The motivation for taking $\hat{g}_{\mu\nu}$ to be a $S^4$ metric in Section \ref{s2} is that only in the case of maximal symmetry it is easy to identify the decoupling scale in terms of geometric data.}.
(It is also interesting that with the stalk removed the computer-generated universes \cite{agjl} are almost perfect $4$-spheres $S^4$.)  

\section{Summary}\label{s4}

We discussed the effective potential of the conformal factor both from a QEG and a CDT perspective. We demonstrated that if QEG is asymptotically safe then it gives rise to a potential which becomes flat for $\phi \rightarrow 0$, allowing for a phase of gravity with vanishing metric expectation value. The argument assumes the existence of an underlying UV fixed point, but is exact otherwise. We argued that the potential $V (a)$ ``measured'' in Monte Carlo simulations within the CDT approach does indeed reflect the predicted behavior at small scale (or conformal) factors. It is therefore plausible to assume that already the presently available CDT simulations, indirectly, have seen evidence for the same NGFP that has been found in QEG by truncated flow equations. This hints at the possibility that QEG and CDT might be related at a deeper level.
\\
\\ 
\\
Acknowledgement: M.R. would like to thank J.~Ambj\o{}rn, H.~Hamber, R.~Loll, and R.~Williams for helpful discussions.

%
%
%
%\subsection{}\label{s3.1}
%
% \eqref{2.crehtrunc} but instead of the single quartic term 

%
%

%

%
%
%
%%%%%%%%%%%%%%%%%%%%%%%%%%%%%%%%%%%%%%%%%%%%%%%%%%%%%%%%%%%%
%%%%%%%%%%%%%%%%%%%%%%%%%%%%%%%%%%%%%%%%%%%%%%%%%%%%%%%%%%%%
%
%
%
%
%%%%%%%%%%%%%%%%%%%%%%%%%%%%%%%%%%%%%%%%%%%%%%%%%%%%%%%%%%%%
%%%%%%%%%%%%%%%%%%%%%%%%%%%%%%%%%%%%%%%%%%%%%%%%%%%%%%%%%%%%
\pagebreak

\end{document}